\documentclass[useAMS,usenatbib]{mnras}

\usepackage{times}
\usepackage{amssymb, amsmath}
\usepackage{graphicx,rotating,amssymb,longtable,lscape}
\usepackage{epsfig}
\usepackage{enumerate}
\usepackage{hyperref}

\def\apj{\mbox{ApJ}}
\def\apjl{\mbox{ApJL}}
\def\apjs{\mbox{ApJS}}

\def\mnras{\mbox{MNRAS}}
\def\aj{\mbox{AJ}}
\def\araa{\mbox{ARA\&A}}
\def\pasp{\mbox{PASP}}
\def\nat{\mbox{Nature}}
\def\aap{\mbox{A\&A}}

\title[Dissecting the AGN in Circinus]{Dissecting the active galactic nucleus in Circinus -- I. Peculiar mid-IR morphology explained by a dusty hollow cone\thanks{Based on European Southern Observatory (ESO) observing programmes 60.A-9629, 076.B-0599, 078.B-0255, 087.B-0746, 088.B-0159, 089.B-0070, 385.B-0896, and 386.B-0026}
}

\author[Stalevski, Asmus \& Tristram]{Marko Stalevski$^{1,2,3}$\thanks{E-mail: marko.stalevski@gmail.com}, Daniel Asmus$^{4}$ and Konrad R. W. Tristram$^{4}$
\\
$^{1}$Astronomical Observatory, Volgina 7, 11060 Belgrade, Serbia\\
$^{2}$Departamento de Astronom\'\i a, Universidad de Chile, Camino El Observatorio 1515, Casilla 36-D Santiago, Chile\\
$^{3}$Sterrenkundig Observatorium, Universiteit Gent, Krijgslaan 281-S9, Gent, 9000, Belgium\\
$^{4}$European Souther Observatory, Casilla 19001, Santiago 19, Chile}

\begin{document}

\date{\today}

\pagerange{\pageref{firstpage}--\pageref{lastpage}} \pubyear{2017}

\maketitle

\label{firstpage}

\begin{abstract}
Recent high angular resolution observations resolved for the first time the mid-infrared (MIR) structure of nearby active galactic nuclei (AGN). Surprisingly, they revealed that a major fraction of their MIR emission comes from the polar regions. This is at odds with the expectation based on AGN unification, which postulates a dusty torus in the equatorial region. The nearby, archetypical AGN in the Circinus galaxy offers one of the best opportunities to study the MIR emission in greater detail. New, high quality MIR images obtained with the upgraded VISIR instrument at the Very Large Telescope show that the previously detected bar-like structure extends up to at least $40\,$pc on both sides of the nucleus along the edges of the ionization cone. Motivated by observations across a wide wavelength range and on different spatial scales, we propose a phenomenological dust emission model for the AGN in the Circinus galaxy consisting of a compact dusty disk and a large-scale dusty cone shell, illuminated by a tilted accretion disk with an anisotropic emission pattern. Undertaking detailed radiative transfer simulations, we demonstrate that such a model is able to explain the peculiar MIR morphology and account for the entire IR spectral energy distribution. Our results call for caution when attributing dust emission of unresolved sources entirely to the torus and warrant further investigation of the MIR emission in the polar regions of AGN.
\end{abstract}

\begin{keywords}
galaxies: active -- galaxies: nuclei -- galaxies: Seyfert -- radiative transfer.
\end{keywords}

\section{Introduction}
\label{sec:intro}

During the lifetime of a galactic nucleus, the supermassive black hole (SMBH) in its center may accrete significant amounts of matter at a relatively high rate. This phenomenon, know as an active galactic nucleus (AGN), manifests itself through a number of energetic phenomena such as: a compact X-ray source, a strong UV$/$optical continuum, a number of broad and$/$or narrow emission lines, a radio-jet, to name a few. Approximately half of the luminosity radiated by the matter spiraling onto the SMBH in an accretion disk is absorbed by the surrounding dust and reemited in the infrared (IR). The distribution of dust and the viewing angle determine how an AGN appears to an observer as they combine to reveal or hide the accretion disk and broad line region \citep[AGN type 1 and type 2, respectively; see][]{Antonucci1993,Urry-Padovani1995}. This dusty material, postulated to be found preferentially in the equatorial plane of the system and roughly in a toroidal shape, has been rooted in AGN jargon as ``the dusty torus'', even though a number of both theoretical and observational pieces of evidence suggest that it is more likely to resemble a complex multiphase medium, possibly associated to dusty outflows or failed winds. A bulk of observational evidence has been accumulated over the years in support of this scenario \citep[][and references therein]{Netzer2015}.

Owing to their small angular sizes, only with recent advances of IR interferometry several nearby sources could be resolved, revealing unexpected and surprising results. Namely, it has been found that the mid-infrared (MIR) emission of these sources appears elongated in the polar direction, perpendicular to the plane of the dusty torus. This was firmly established for the two nearest, archetypal Seyfert (Sy) 2 galaxies, NGC 1068 \citep{Jaffe2004, Wittkowski2004, Raban2009, Lopez-Gonzaga2014} and the Circinus galaxy \citep[hereafter referred to as Circinus; ][]{Tristram2007,Tristram2014}. Both of them were decomposed into a parsec-sized disk and a component extended in the polar direction. Similar extensions of the thermal dust emission have been confirmed for another Sy 2, NGC 424 \citep{Honig2012}, and an inclined Sy 1, NGC 3783 \citep{Honig2013}.
\citet{Lopez-Gonzaga2016} analyzed the sample of 23 AGN observed by MIR interferometry from \citet{Burtscher2013} and found that seven sources have sufficiently good $(u,v)$ coverage and signal-to-noise (S$/$N) to detect extended emission, if present. Of these seven sources, five are confirmed to have polar extension (the four already mentioned above and one new detection, NGC 5506). Polar elongation of the MIR emission in NGC 1068 was reported even in single dish data \citep{Braatz1993, Bock2000}. More recently, \citet{Asmus2016} established 18 objects with extended MIR polar emission in single dish data on the scales of tens to hundreds of parsecs. The polar extension is found in virtually all sources with sufficient S$/$N data and favorable viewing angle (close to edge-on). Furthermore, the orientation of the MIR polar extension on these large scales matches the orientation of the polar component on parsec-scales revealed by interferometry, indicating that both might have the same physical origin, possibly in the dusty winds driven by radiation pressure on the dust grains in the innermost region. In this picture, what is now called ``the torus'' might actually be a puffed-up inner rim of a dusty disk from which the dust is lifted up (see \S~\ref{sec:origin}).

In all cases discussed above the extended polar component represents a significant or even a major fraction of the total IR flux. These findings challenge the use of standard dusty torus models to interpret the IR emission and demand a new paradigm for the dust structure in AGN. Crucial steps towards the new paradigm are case studies of nearby objects that can be resolved and studied extensively across a broad range of wavelengths. One such object is the archetypal type 2 AGN at the distance of $4.2\,$Mpc in the Circinus galaxy. The galaxy is inclined by $65^{\circ}$ and the nucleus is heavily obscured by dust lanes in the plane of the galaxy \citep{Prieto2004, Mezcua2016}. Being the second brightest AGN in the MIR and allowing high intrinsic spatial resolution ($1\arcsec=20\,$pc), Circinus is a prime target for a large number of studies across a wide wavelength range. These observations revealed a number of features expected from a prototypical type 2 AGN: narrow emission lines, a Compton-thick nucleus \citep{Ricci2015}, a prominent Fe K$_{\alpha}$ emission line \citep{Arevalo2014} and a prominent ionization cone. [\ion{O}{iii}] and H$_{\alpha}$ emission reveal the ionization cone on the West side \citep{Marconi1994, Veilleux1997, Wilson2000}; the East cone is covered by the host galaxy disk but was detected in polarized near-infrared light \citep{Ruiz2000} and the infrared coronal [\ion{Si}{vii}] $2.48\,\micron$ emission line \citep{Prieto2004}. Water maser emission traces a near-Keplerian, warped disk seen edge-on and a parsec-scale outflow corresponding well to the orientation and opening angle of the ionization cone \citep{Greenhill2003}. Evidence of outflows is seen also in [\ion{O}{iii}] and CO. \citet{Packham2005} detected and resolved the bar-like extended emission at $8.7$ and $18.3\,\micron$ for the first time, extending up to $\sim2\arcsec$ from each side of the nucleus, roughly in east-west direction. Modeling of the MIR interferometric data obtained with the MID-infrared Interferometric instrument (MIDI) at the Very Large Telescope Interferometer (VLTI) reveals two components on parsec-scale: a disk-like component in the equatorial plane of the system, and a larger structure elongated in the polar direction \citep{Tristram2007, Tristram2014}. The disk-like component is likely a molecular, dusty extension of the accretion disk. Furthermore, it well matches the orientation and scale of the warped maser disk. The association of the polar-extended component is less clear. It could be the inner wall of the torus, which then would need to have a very large scale height, or it could be the base of a polar dusty wind, which is forming a large cone shell.

The upgraded Imager and Spectrometer for mid-Infrared (VISIR; \citealt{lagage_successful_2004}) mounted on the Very Large Telescope (VLT) provided up-to-date highest quality MIR images of Circinus. These images show a previously detected, prominent bar extending $40\,$pc on both sides of the unresolved nucleus. This bar cannot be explained by the torus: it is aligned with the polar component inferred by interferometric data and with the edge of the ionization cone seen at optical wavelengths. 

In this work we propose a model for dust emitting regions of AGN in Circinus that may explain this puzzling finding. We start by presenting the new data which motivated this work in \S \ref{sec:obs}. In \S \ref{sec:mod}, we describe the geometry and dust properties of our model. Results of the radiative transfer simulations of the models and comparison with observations are given in \S \ref{sec:res}. We summarize and lay out our conclusions in \S \ref{sec:sum}.

\section{Observational data}
\label{sec:obs}

In this section, we first describe the new high-fidelity $N$-band images of Circinus obtained with VLT/VISIR which are used for the detailed, morphological comparison with the models. Then, we summarize the additional data used for the comparisons of the spectral energy distribution (SED), including so far unpublished archival $L$-band data from the Infrared Spectrometer And Array Camera \citep[ISAAC,][]{Moorwood1998}, also mounted on the VLT.

\subsection{Data for the morphological comparison}
\label{sec:obs-vlt}

We selected the highest quality MIR data available for Circinus. In particular, two new images in the PAH1 ($8.6\,\micron$) and PAH2\_2 filters ($11.9\,\micron$) were obtained with VISIR as part of its science verification in February and March 2016 (Program ID 60.A-9629) after its upgrade \citep{Kerber2016, Asmus2016_ver}. The images were recorded with standard chopping and nodding in burst mode with a total exposure time of 23\,min each, and two calibrator stars from \citet{Cohen1999} were observed consecutively with Circinus, HD\,119193 before and HD\,128068 after.
The data were custom reduced using the shift \& add technique for the individual exposures of 25~ms (PAH1) and 21~ms (PAH2\_2), allowing us to achieve diffraction-limited image quality in the reduced images.

Since both the PAH1\footnote{\label{ftn:filters} Polycyclic aromatic hydrocarbon (PAH) features and [\ion{S}{iv}] ($10.4\,\micron$) are absent or very weak in the central 100\,pc of Circinus and thus, despite the filter names, all the VISIR filters used here predominantly trace the continuum emission, with insignificant contamination by PAH or coronal line emission (c.f. the Spitzer spectrum in Fig. 8).} and PAH2\_2\textsuperscript{\ref{ftn:filters}} filters trace continuum, we complement them with archival VISIR images in the filters SIV\textsuperscript{\ref{ftn:filters}} ($10.5\,\micron$) and Q2 ($18.7\,\micron$) tracing the two silicate features at $\sim 10$ and $\sim 18\,\micron$. 
These data were taken as part of the programs 076.B-0599 and 078.B-0255 in 2006 in standard imaging mode with exposure times of 12 and 17\,min respectively. Here, we simply take the reduced data from \citet{Asmus2014} and refer the reader to that work for more details on these data.

\begin{figure*}
\centering
\includegraphics[width=1\textwidth]{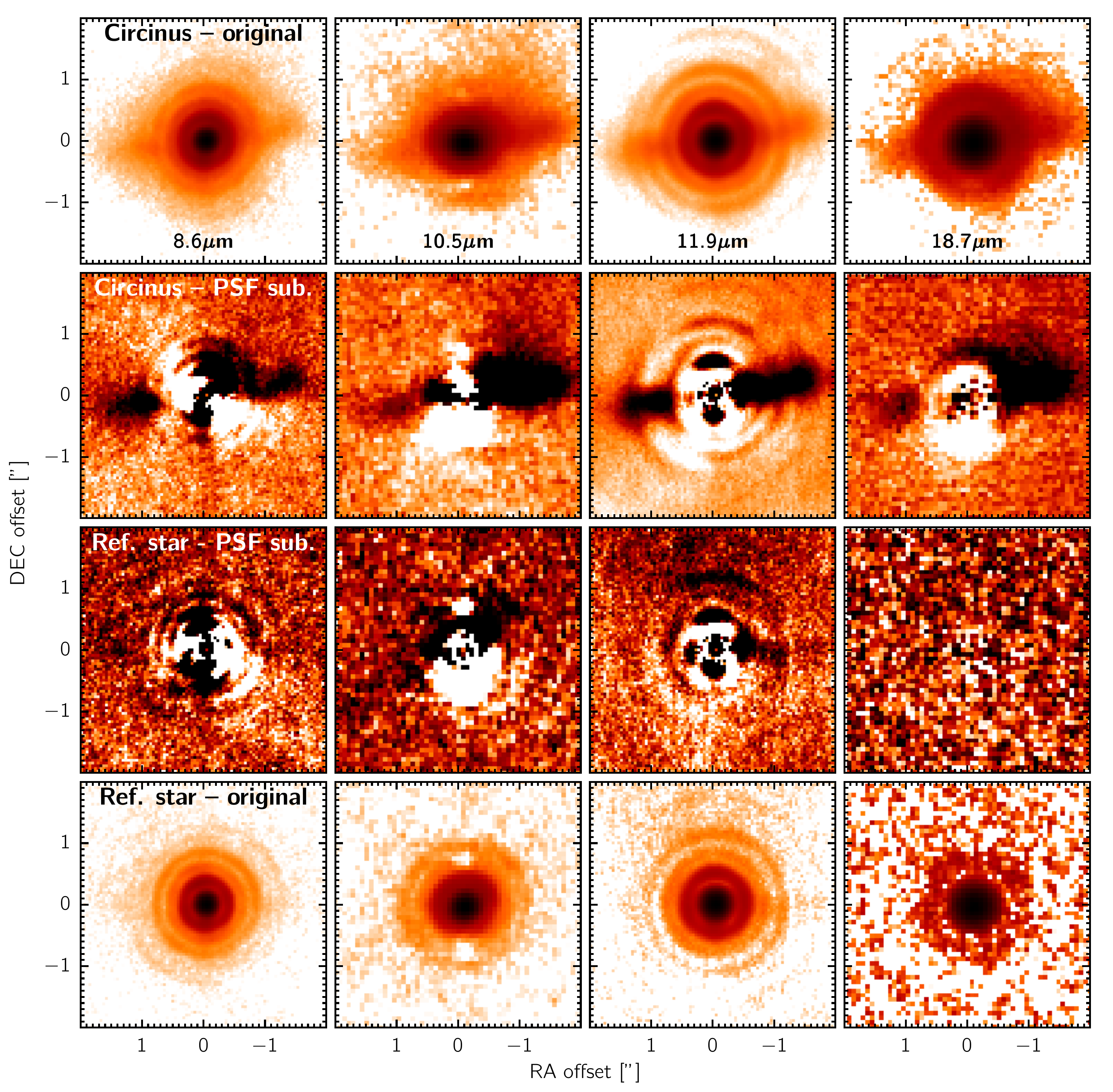}
\caption{Best available MIR images of the AGN in Circinus acquired with VLT/VISIR (top row) and the corresponding reference star (bottom row) in different filters. The color scaling in the top and bottoms rows is logarithmic with white corresponding to the median background levels and black to the maximum count levels. The second and third rows show PSF subtracted images of Circinus and the reference star, respectively, in linear scaling with minimum (white) and maximum (black) corresponding to the 5 and 95\% percentiles. In all images North is up and East to the left.}
\label{fig:cirVLT}
\end{figure*}

The resulting images of the central $4\times4\arcsec$ region of the Circinus galaxy are shown in the top row of Fig.~\ref{fig:cirVLT}.
A bright, unresolved nucleus dominates the emission in all images.
Only with logarithmic colour scaling (as used in Fig.~\ref{fig:cirVLT}), resolved emission in the form of a bar stretching out from the center at a position angle of roughly $100^{\circ}$ both to the East and West is revealed.
This emission structure is well visible at all four wavelengths out to $\sim 40\,$pc on both sides, and coincides with the edge of the ionization cone seen at visible wavelengths on the Western side. At $10.5\,\micron$ and $18.7\,\micron$, the bar is weaker on the Eastern side, while additional emission is visible on the West, smoothly extending to the North from the bar. 
This indicates stronger silicate absorption in the East with respect to the West, likely due to the extinction by the inclined host galaxy disk, in agreement to what \cite{Roche2006} already found.
The larger number of Airy rings in the newer ($8.6\,\micron$ and $11.9\,\micron$) images compared to the older ones showcases the significantly higher S$/$N of the new data.
In addition, the $18.7\,\micron$ image has a particularly low S$/$N, owing to the decreasing atmospheric transmission at these wavelengths. 

To make the extended emission better visible, we subtract the unresolved nucleus represented by the point spread function (PSF) in the second row of Fig.~\ref{fig:cirVLT}.
The extended emission features become very prominent even in the linear colour scale used in those images.
Here, the PSF is approximated by the azimuthally median-averaged PSF, which turns out to be superior to obscured Airy disk model fits to the PSF of the VLT. 
The bottom row of Fig.~\ref{fig:cirVLT} shows the corresponding standard star observations for each of the filters, normally used as PSF reference.
The third row shows the same PSF subtraction method applied to these reference star images and illustrates the residuals this technique leaves.

\subsection{Data for SED comparison}
\label{sec:obs-lit}

For comparison of the model and the data over a wider wavelength range, we also assemble the observed nuclear SED and spectra of Circinus from data available in the literature \citep{Quillen2001, Prieto2004, Packham2005, Asmus2014, Tristram2014}. In addition to already published values, we took all the available subarcsecond-resolution MIR images from \citet{Asmus2014} and performed $4\arcsec$ aperture photometry on them to capture the total flux of the system. For data from multiple epochs between 2010 and 2012 we use averaged measurements. For ground-based observations we assume the typical 10\% uncertainty related to the flux calibration \citep{Cohen1999}; if the measurement error is larger, it is quoted instead. The aperture of the Spitzer/IRS spectrum \citep{Asmus2014} is comparable to the total aperture of VISIR in the $5.2-14.5\,\micron$ range, while significantly larger at longer wavelengths. The MIDI spectrum was extracted using a $\sim0.54\arcsec\times0.52\arcsec$ aperture and hence corresponds to the unresolved nucleus with VISIR \citep{Tristram2014}. Photometric data and all the relevant information are summarized in the table \ref{tab:data}.

Finally, we also include so far unpublished $L$-band data\footnote{Programmes 385.B-0896(B), 386.B-0026(B), 087.B-0746(B), 088.B-0159(B), 089.B-0070(B)} from VLT/ISAAC to obtain a contemporary estimate for the 3.8\,$\mu$m flux. We applied standard chop-nodding subtraction using a custom IDL script. 
Photometry was extracted (using {\tt apphot}) in a $0.4$ and a $4.0\arcsec$ aperture with the sky annulus between $6$ and $7\arcsec$. For the final average flux, we only used those 9 measurements with good image quality (taken between 2010 and 2012), where the full-width-half-maximum ($\mathrm{FWHM}$) $<0.51\arcsec$, yielding $F_{\mathrm{L}}\left(0.4\arcsec\right)=0.98\pm0.13\,\mathrm{Jy}$ and $F_{\mathrm{L}}\left(4.0\arcsec\right)=1.21\pm0.08\,\mathrm{Jy}$ for the two apertures respectively. In all cases, HD\,129858 was used as the calibrator star. The uncertainties in the final measurement reflect the scatter of the individual measurements. No trends in the flux values which could be caused by variability were detected over the measurement period, i.e. the Circinus nucleus was in a relative stable state during the above cited period of measurements. In our modeling, we use only $L$-band photometry, since the images are of relatively low quality and extended emission is dominated by the host galaxy.

\begin{table*}
\centering
\caption{Photometric data of Circinus compiled from the literature and presented in this work.}
\label{tab:data}
\begin{tabular}{llllccccc}
\hline\hline
band            & instrument & year      & reference             & $\lambda\,[\micron]$ & HW $[\micron]$ & aper. $[\arcsec]$ & $\lambda F_{\lambda}$ $\mathrm{[W/m^2]}$ & $\varDelta\lambda F_{\lambda}$ $\mathrm{[W/m^2]}$ \\
(1)             & (2)        & (3)       & (4)                   & (5)                  & (6)            & (7)               & (8)                     & (9) \\
\hline
J               & NACO       & 2003      & \citet{Prieto2004}    & 1.3                  & 0.1               & 0.38                 & 3.7E-15        & 3.7E-16       \\
H               & NICMOS     & 1998      & \citet{Quillen2001}   & 1.7                  & 0.2               & 0.10                 & 8.5E-15        & 1.2E-15       \\
K               & NACO       & 2003      & \citet{Prieto2004}    & 2.2                  & 0.2               & 0.38                 & 2.6E-14        & 2.6E-15       \\
$2.48\,\micron$ & NACO       & 2003      & \citet{Prieto2004}    & 2.4                  & 0.0               & 0.38                 & 3.8E-14        & 3.8E-15       \\
L               & NACO       & 2003      & \citet{Prieto2004}    & 3.8                  & 0.3               & 0.38                 & 3.0E-13        & 3.0E-14       \\
L               & ISAAC      & 2010-2012 & this work             & 3.8                  & 0.3               & 0.40                 & 7.7E-13        & 1.0E-13       \\
L               & ISAAC      & 2010-2012 & this work             & 3.8                  & 0.3               & 4.00                 & 9.5E-13        & 6.3E-14       \\
M               & NACO       & 2003      & \citet{Prieto2004}    & 4.5                  & 0.3               & 0.38                 & 1.3E-12        & 1.3E-13       \\
PAH1            & VISIR      & 2010-2011 & \citet{Asmus2014}     & 8.6                  & 0.2               & 0.40                 & 1.7E-12        & 2.8E-13       \\
PAH1            & VISIR      & 2016      & this work             & 8.6                  & 0.2               & 0.40                 & 1.0E-12        & 1.0E-13       \\
PAH1            & VISIR      & 2010-2011 & this work             & 8.6                  & 0.2               & 4.00                 & 3.3E-12        & 3.6E-13       \\
PAH1            & VISIR      & 2016      & this work             & 8.6                  & 0.2               & 4.00                 & 3.4E-12        & 3.4E-13       \\
Si2             & T-ReCS      & 2004      & \citet{Asmus2014}     & 8.7                  & 0.2               & 0.40                 & 2.2E-12        & 2.2E-13       \\
Si2             & T-ReCS      & 2004      & this work             & 8.7                  & 0.2               & 4.00                 & 3.2E-12        & 3.2E-13       \\
SIV             & VISIR      & 2006      & \citet{Asmus2014}     & 10.5                 & 0.1               & 0.40                 & 9.3E-13        & 9.3E-14       \\
SIV             & VISIR      & 2006      & this work             & 10.5                 & 0.1               & 4.00                 & 1.8E-12        & 1.8E-13       \\
SIV\_2          & VISIR      & 2006      & \citet{Asmus2014}     & 10.8                 & 0.1               & 0.40                 & 1.0E-12        & 1.0E-13       \\
SIV\_2          & VISIR      & 2006      & this work             & 10.8                 & 0.1               & 4.00                 & 2.2E-12        & 2.2E-13       \\
PAH2\_2         & VISIR      & 2006      & \citet{Asmus2014}     & 11.9                 & 0.2               & 0.40                 & 2.0E-12        & 2.0E-13       \\
PAH2\_2         & VISIR      & 2010-2011 & \citet{Asmus2014}     & 11.9                 & 0.2               & 0.40                 & 2.4E-12        & 2.9E-13       \\
PAH2\_2         & VISIR      & 2016      & this work             & 11.9                 & 0.2               & 0.40                 & 3.1E-12        & 3.1E-13       \\
PAH2\_2         & VISIR      & 2006      & this work             & 11.9                 & 0.2               & 4.00                 & 3.2E-12        & 3.2E-13       \\
PAH2\_2         & VISIR      & 2010-2011 & this work             & 11.9                 & 0.2               & 4.00                 & 4.3E-12        & 3.3E-13       \\
PAH2\_2         & VISIR      & 2016      & this work             & 11.9                 & 0.2               & 4.00                 & 4.5E-12        & 4.5E-13       \\
Q1              & VISIR      & 2008      & \citet{Asmus2014}     & 17.7                 & 0.4               & 0.60                 & 3.2E-12        & 3.2E-13       \\
Q1              & VISIR      & 2008      & this work             & 17.7                 & 0.4               & 4.00                 & 4.6E-12        & 4.6E-13       \\
Qa              & T-ReCS      & 2004      & \citet{Asmus2014}     & 18.3                 & 0.4               & 0.60                 & 2.2E-12        & 2.2E-13       \\
Qa              & T-ReCS      & 2004      & this work             & 18.3                 & 0.4               & 4.00                 & 2.9E-12        & 2.9E-13       \\
Q2              & VISIR      & 2006      & \citet{Asmus2014}     & 18.7                 & 0.4               & 0.60                 & 2.5E-12        & 2.5E-13       \\
Q2              & VISIR      & 2006      & this work             & 18.7                 & 0.4               & 4.00                 & 3.9E-12        & 3.9E-13       \\
Q3              & VISIR      & 2008      & \citet{Asmus2014}     & 19.5                 & 0.2               & 0.60                 & 2.9E-12        & 2.9E-13       \\
Q3              & VISIR      & 2008      & this work             & 19.5                 & 0.2               & 4.00                 & 4.6E-12        & 4.6E-13      \\
\hline\hline
\end{tabular}
\begin{minipage}{1.0\textwidth}
 
{\it -- Notes:} 
(1) Filter band;
(2) instrument;
(3) year the data are taken;
(4) reference;
(5) central wavelength of the filter;
(6) filter band half width;
(7) aperture;
(8) flux;
(9) measurement uncertainty.
        
\end{minipage}
\end{table*}

\section{The model of the Circinus AGN}
\label{sec:mod}

In this section we provide a detailed description of our model including geometry, dust properties, adopted parameter values and method employed for producing model images and SEDs.

\subsection{Model geometry}
\label{sec:mod-geo}

\begin{figure*}
\centering
\includegraphics[width=1.0\textwidth]{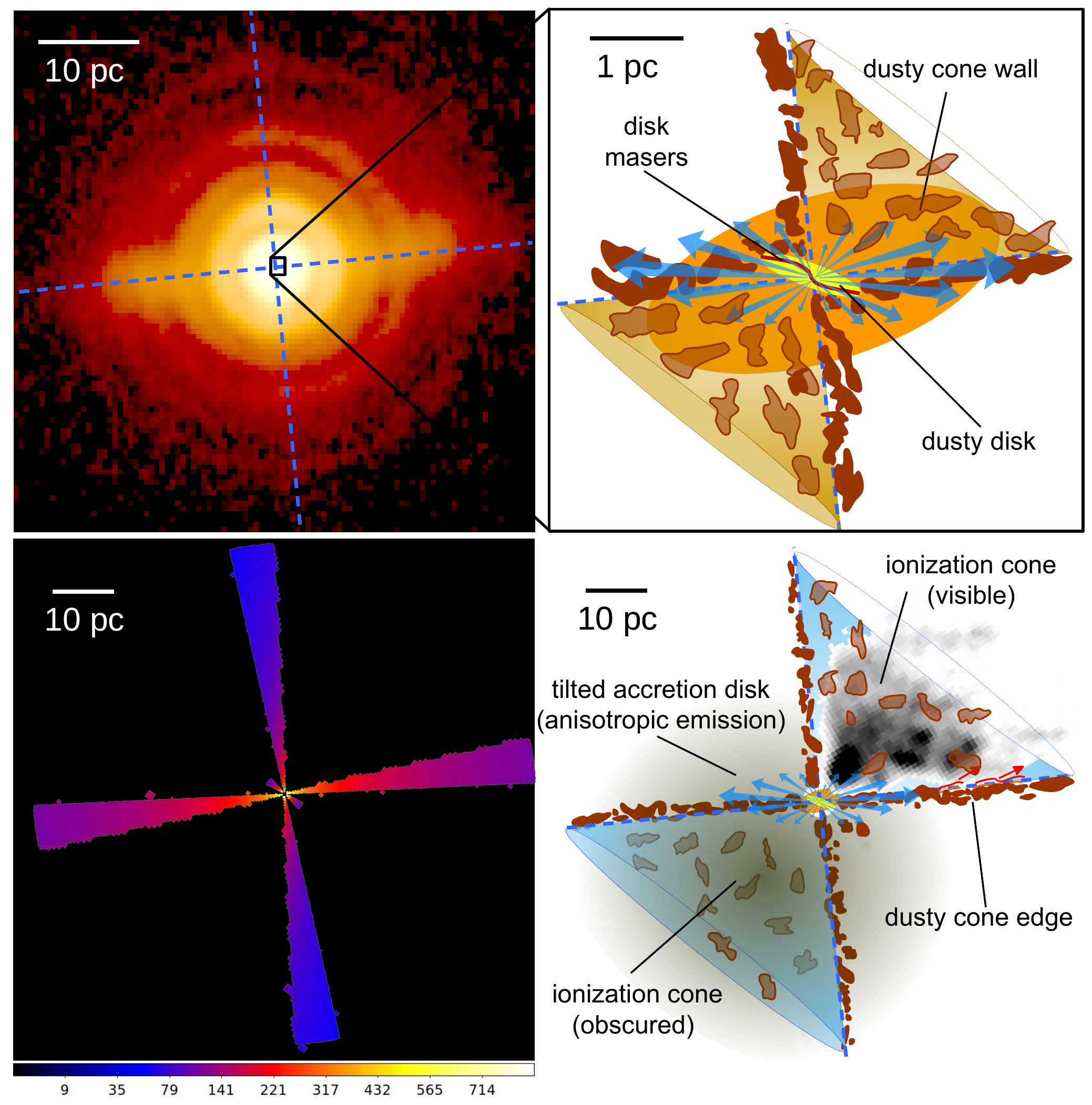}
\caption{Schematic of the model proposed here for the nuclear dust distribution in Circinus consisting of a compact dusty disk and a large scale dusty cone, compared with observations at different wavelengths and scales. \emph{Upper left panel}: $12\,\micron$ VLT/VISIR image revealing the prominent bar. \emph{Upper right panel:} inner region on scales of a few parsecs resolved by VLT/MIDI into the disk and polar components (yellow and orange ellipsoids). The red line within the yellow disk component traces the warped disk in water maser emission \citep{Greenhill2003}. \emph{Lower right panel}: The large scale structure, with the optical ionization cone image (from \citealt{Wilson2000}; in gray scale) overlaid with the model schematic. The blue arrows illustrate the anisotropic emission pattern of the accretion disk, whose orientation matches the orientation of the inner part of the warped maser disk. Dashed blue lines in all the panels are tracing the edges of the ionization cone. \emph{Lower left panel}: A slice of the temperature map in the vertical plane of the disk$+$cone model obtained from Monte Carlo radiative transfer simulations.}
\label{fig:scheme}
\end{figure*}

The large scale MIR bar (extending up to $40\,$pc on both sides of the nucleus) seen by VISIR is aligned with the polar-elongated component on parsec-scale, suggesting that both are part of the same physical structure. The big surprise is that this structure is dominating the MIR emission: $\sim10\%$ of the total flux in the MIR is coming from the disk-like component ($\sim0.2-1.1\,$pc), about 40-50\% from the parsec-scale polar component ($\sim0.8-1.9\,$pc), and another 40-50\% from the large-scale polar bar \citep{Asmus2016}. The quoted sizes correspond to the FWHM of the black-body emitters with a Gaussian brightness distribution, used to model the interferometric data by \citet{Tristram2014}. On the other hand, if the large-scale polar bar is part of a hollow dusty cone, one would expect this cone to appear as an X-shaped structure. 
Hence the question arises: why the other side of the cone wall does not appear in the VISIR images? \emph{Our hypothesis is that the inner accretion disk is significantly tilted (and/or warped) so that it preferentially illuminates only one side of the dusty cone wall.} This picture is supported by the orientation of the inner part of the warped water maser disk and by the anisotropic illumination pattern imprinted in the ionization cone.

In Fig.~{\ref{fig:scheme}}, we present a schematic of our model that has the potential to explain the MIR bar seen with VISIR, and at the same time fits well into the picture placed by the previous observations at different wavelengths, both on small and large scales. The overall geometry is well constrained by observations: a parsec-scale dusty disk, co-spatial with the warped maser disk seen edge-on (upper right panel), and an ionization cone seen in the optical, extending out to $\sim40\,$pc from the nucleus (lower right panel). The illumination pattern of the ionization cone (brighter toward the western edge) is indicative of the anisotropic emission pattern of the ionizing source. This is corroborated by the orientation of the inner part of the maser disk, which is roughly perpendicular to the cone edge. An optically-thick, geometrically-thin accretion disk displays a $\cos\theta$ angular-dependent luminosity profile \citep{Netzer1987}. 
Aligned with the inner part of the warped maser disk, such a disk will emit more strongly into or close to the western edge of the cone. If the cone wall is dusty, then the described setup could naturally produce the dusty bar seen in the VISIR image. The opposite side of the cone wall would remain cold and invisible, as the tilted anisotropic accretion disk would emit very little in that direction (see temperature profile in the lower left panel).

\begin{figure*}
\centering
\includegraphics[width=1.0\textwidth]{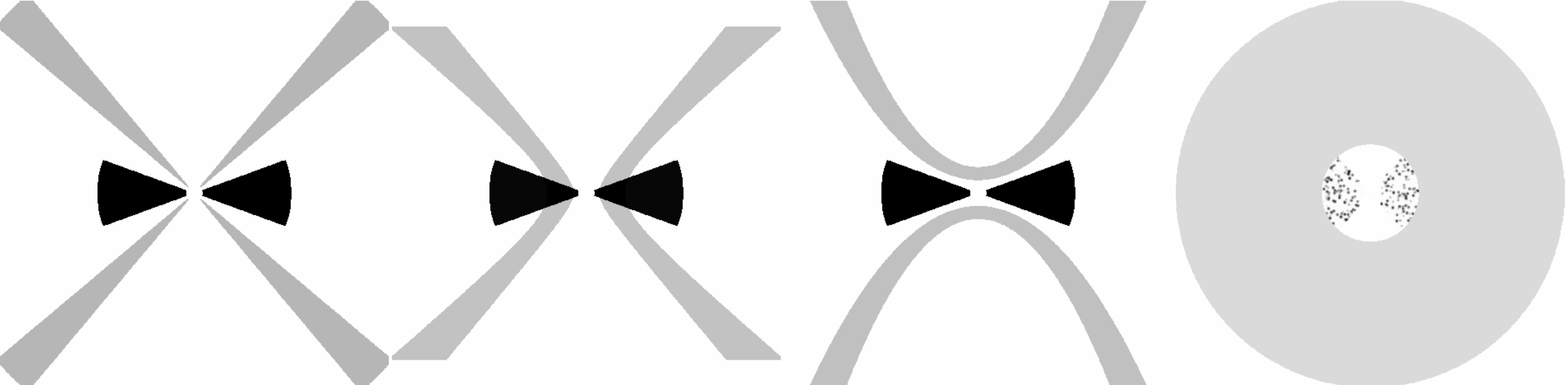}
\caption{Schematic representation of the tested model geometries, from left to right: disk$+$cone shell, disk$+$hyperboloid shell, disk$+$paraboloid shell, and a clumpy torus surrounded by a dusty sphere shell. For clarity of the illustration, the sizes of the different components are exaggerated and not on scale with respect to each other.}
\label{fig:geoscheme}
\end{figure*}

Apart from a cone geometry, which has perfectly straight walls, two other geometries of the dust in polar region might produce similar results: a one-sheeted hyperboloid shell and a paraboloid shell, whose walls are obtained by rotating a hyperbola and a parabola, respectively. In all cases, the geometries are characterized by the half opening angles of the inner and outer walls of the shell, the radial extension, and the inner radius (or offset from the system center). We also consider a case of a canonical clumpy torus surrounded by a dusty sphere shell in order to test if simply radiation collimated by the torus and illuminating diffuse dusty material surrounding an AGN can result in the observed MIR morphology. All described geometries are sketched in Fig.~{\ref{fig:geoscheme}}.

\subsection{Model parameters}
\label{sec:mod-par}

The starting model consists of two components whose parameters take fiducial values implied by the different observations discussed in \S \ref{sec:intro}: (a) a compact flared dusty disk with a flaring angle of $\Delta=20^{\circ}$ (measured from the equatorial plane to the disk edge) and an outer radius of $R_{\text{out}}=1.5\,$pc, as well as (b) a hollow dusty cone with the full angular width of $10^{\circ}$ extending out to $40\,$pc and a half opening angle of 40$^{\circ}$ (measured from the system polar axis). The dust is distributed according to the following density law:
\begin{equation}\label{eqn:rho}
\rho\left(r,\theta \right)\propto r^{-p}e^{-q|\cos\theta|} ,
\end{equation}
where $r$ and $\theta$ are coordinates in the spherical coordinate system.
For the starting model, both components possess a constant density profile ($p=q=0$). The disk is optically thick in the MIR (edge-on optical depth of $\tau_{9.7}=5$), while the dust mass of the cone is chosen so that the line of sight passing all the way through the cone shell is at the limit of being optically thick in the $V$-band ($\tau_{V}=1$). The optical depth along the cone wall cannot be much higher as most of the optical/UV emission would then be absorbed and reprocessed in the inner regions, and could not produce the observed MIR morphology, that is the large scale bar. It cannot be much lower either: while more light would be able to escape further out, there would not be enough dust, and the outer regions would remain too faint in the MIR. The accretion disk is tilted by $40^{\circ}$ with respect to the system polar axis. Everything said above for the cone model applies as well to the hyperboloid and paraboloid models; the only difference being the shape of their walls.

The dust in the above models is distributed homogeneously. A more realistic representation would likely be a clumpy distribution. However, to allow the light to penetrate far enough, the cone shell would need to have an extremely low filling factor, consisting of a very large number of very small clumps. Given the size of the simulation box ($80\times80\times80\,$pc), resolving such small clumps requires excessive computational resources which would restrict our investigation. However, we confirmed that this assumption does not have a significant impact on the results by comparing the results of a small scale smooth and clumpy test model. 

The last model geometry we consider corresponds to the canonical case of a clumpy ``torus''. It consists of several thousands of clumps distributed inside a flared disk geometry with $\Delta=50^{\circ}$, $\tau_{9.7}=5$ and $R_{\text{out}}=1.5\,$pc, surrounded by a dusty sphere shell extending out to $40\,$pc with a radial optical depth of $\tau_{V}=1$ \cite[for more details see][]{Stalevski2016}.

The adopted accretion disk SED is described by a standard ``blue bump'' composition of power-laws with an anisotropic emission pattern that accounts for the change in the projected surface area and for the limb darkening effect (see equations 2 and 3 in \citealp{Stalevski2016}). Estimates from the literature for the bolometric luminosity of the accretion disk of Circinus are in the range of $L_{\text{AGN}}=1-7\times10^{10} L_{\odot}$ \citep{Moorwood1996, Oliva1999, Tristram2007}. The uncertainty on $L_{\text{AGN}}$ may be even larger as an increase in flux level by a factor of 2 was observed between 2008 and 2010 \citep{Tristram2014}. Here we adopt a value of $L_{\text{AGN}}=4\times10^{10} L_{\odot}$.

The parameters of these fiducial, starting models are summarized in the table \ref{tab:param}.
\begin{table}
\centering
\caption{The adopted values of the parameters that describe the different models considered in this work.}
\label{tab:param}
\begin{tabular}{ll}
                             &                                   \\ \hline\hline
\multicolumn{2}{c}{disk parameters}                        \\ \hline
angular half width           & $20^{\circ}$                      \\
outer radius                 & $1.5$ pc                          \\
radial optical depth         & $\taụ_{9.7}=5$                   \\ \hline
\multicolumn{2}{c}{torus  parameters}                      \\ \hline
angular half width           & $50^{\circ}$                      \\
outer radius                 & $1.5$ pc                          \\
optical depth                & $\taụ_{9.7}=5$                   \\
clump filling factor         & 0.15                              \\ \hline
\multicolumn{2}{c}{sphere shell parameters}                      \\ \hline
inner radius, outer radius   & $1.5$ pc, $40$ pc                 \\
radial optical depth         & $\taụ_{V}=1$                     \\ \hline
\multicolumn{2}{c}{cone/hyperboloid/paraboloid parameters} \\ \hline
half opening angle           & $40^{\circ}$                      \\
angular full width           & $10^{\circ}$                      \\
radial extension             & $40$ pc                           \\
radial optical depth         & $\taụ_{V}=1$                     \\ \hline
\multicolumn{2}{c}{accretion disk parameters}                    \\ \hline
bolometric luminosity        & $4\times10^{10} L_{\odot}$        \\
tilt                         & $0^{\circ}, 40^{\circ}$                   \\ \hline\hline
\end{tabular}

{\it -- Notes:} 
See text for the definitions of the parameters.
\end{table}

\subsection{The dust grain size and composition}
\label{sec:mod-dust}

Silicate grains are destroyed when heated to a temperature above $\sim 1200\,$K, leaving only graphite grains that can survive up to $\sim 1900\,$K. Furthermore, smaller grains (of both types) are destroyed at lower temperatures than the larger grains \citep{Draine1984, Draine-Lee1984, Barvainis1987}. This suggests that the chemical composition of the dust exposed to the strong AGN radiation should be dominated by large graphite grains, unlike the composition of the standard ISM dust. Large, micron-size grains are indeed strongly preferred by the observed optical-to-MIR extinction ratio \citep{Shao2017} and fits to AGN spectra \citep{Xie2017}. A number of interferometric and near-IR reverberation measurements have been interpreted as evidence that dust is located at a distance smaller by factor of 2--3 than the sublimation radius of the silicates \citep{Kishimoto2007, Kishimoto2009b, Honig-Kishimoto2010, Kishimoto2011b}. Although an anisotropic accretion disk can account for shorter time lags \citep{Kawaguchi2010}, the inferred sub-unity emissivity also implies emission from large grains \cite[e.g.][]{Honig2013}. In the specific case of Circinus, the silicate feature strength does not change with longer baselines which probe smaller spatial scales \citep{Tristram2014}: this may be an indication that the silicate absorption is coming from larger scales, caused by foreground dust in the host galaxy. 
Indeed, \citet{Goulding2012} established that very deep silicate absorption is associated with either edge-on host galaxies, or with prominent dust lanes and/or disturbed morphologies.

Dust in the polar regions is very likely brought there from the innermost region, close to the sublimation radius, by a wind driven by radiation pressure on the dust grains (see \S~\ref{sec:origin}). It follows that the dust in the cone shell should be dominated by large graphite grains. This readily explains why the $10\,\micron$ silicate features in type 1 AGN are typically weaker than predicted by the emission models, without the need to invoke a particular geometry or radiative transfer effects to attenuate the feature. Since silicates are still present in the outer, colder regions of the dusty disk/torus, the silicate feature would appear in absorption when the system is seen close to edge-on, as observed in type 2 AGN. We do not exclude the presence of silicates in the polar region: they could reform, if the physical conditions in the wind are favorable. However, as discussed in \S~\ref{sec:res-SED}, there is a significant foreground absorption which is dominating the $10\,\micron$ silicate feature. Thus, the silicate emission and foreground absorption are degenerate and cannot be disentangled from the silicate feature profile. Our results remain valid even if a certain amount of silicate is present in the polar region: the model would then simply demand a correspondingly higher level of foreground extinction.

In summary, we adopt the dust composition consisting of a mixture of graphite and silicates in the disk and only graphite in the cone shell. The dust grain size is in the range of $a=0.1-1\,\micron$ with the size distribution following the standard MRN power-law $\propto a^{-3.5}$ \citep*{MRN1977}. 

\subsection{Radiative transfer simulations}
\label{sec:mod-mcrt}

To obtain realistic images and SEDs for the proposed model, we employ \textsc{SKIRT}\footnote{\url{http://www.skirt.ugent.be}}, a state--of--the--art 3D radiative transfer code based on the Monte Carlo technique \citep{Baes2011, Baes-Camps2015, Camps-Baes2015}. \textsc{SKIRT} is being used for a variety of applications, including studies of dusty galaxies \citep{Baes2010, DeLooze2012a, DeGeyter2015, Camps2016}, dust in AGN \citep{Stalevski2012a, Stalevski2012b, Stalevski2016}, molecular clouds \citep{Hendrix2015}, and stellar systems \citep{Deschamps2015}. The code uses the Monte Carlo technique to emulate the relevant physical processes including multiple anisotropic scattering and absorption, and computes the temperature distribution of the dust and the thermal dust re-emission self-consistently. Before the simulation starts, the dust of a desired geometry and density profile is distributed on an adaptive grid of a large number of small cells whose size is chosen according to an algorithm which ensures that the dust is properly sampled \citep{Camps2013, Saftly2013, Saftly2014}. The total AGN luminosity is divided into a large number of photons (typically $10^{6}$ per wavelength bin) within a provided SED shape. The photons are then emitted in the surrounding space following the defined angular luminosity profile. Each photon is followed along its path; when it encounters a dust particle, pseudo-random numbers are drawn from proper probability distribution functions to determine its fate. It can scatter and continue its journey, or get absorbed and increase the temperature of the dust in a given cell. In a second stage, the dust itself is a source of thermal photons, which are followed in the same way. This stage is iterated until convergence is reached. One or more virtual instruments are placed around the system to collect the photons and construct images and SEDs, such as those seen in Fig.~{\ref{fig:modImgSED}}.

\section{Results and discussion}
\label{sec:res}

In this section, we first examine the large scale morphology of the dust emission models with different geometries. We simulate how these models look when observed with VISIR to see which of them feature a morphology compatible with the observed MIR images of the Circinus nuclear region. After selecting the model geometries which are qualitatively consistent with the observed images, we proceed with a detailed comparison to the observed SED. 

\subsection{Model vs. data: MIR morphology}
\label{sec:res-large}

\begin{figure*}
\centering
\includegraphics[height=0.9\textwidth]{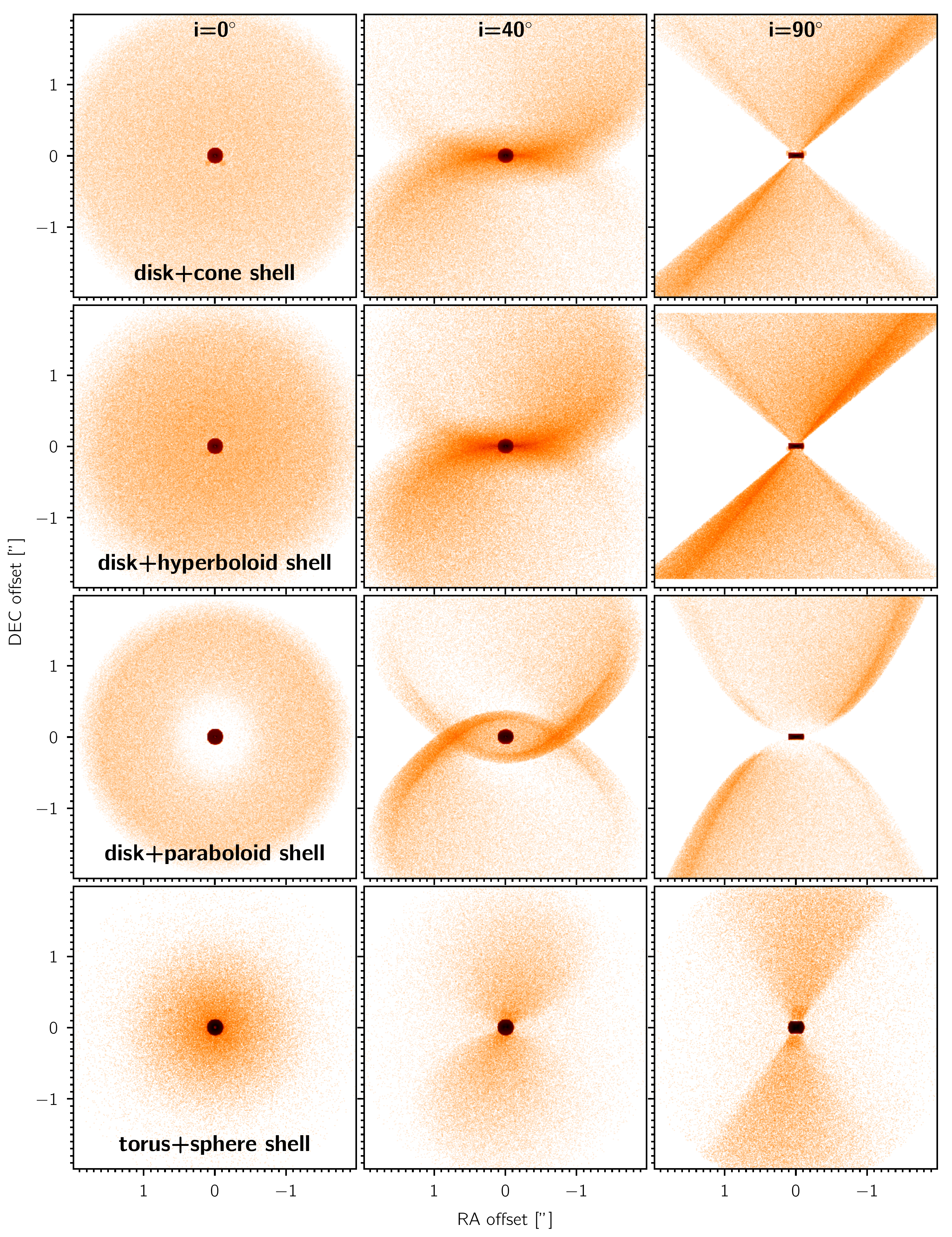}
\includegraphics[height=0.9175\textwidth]{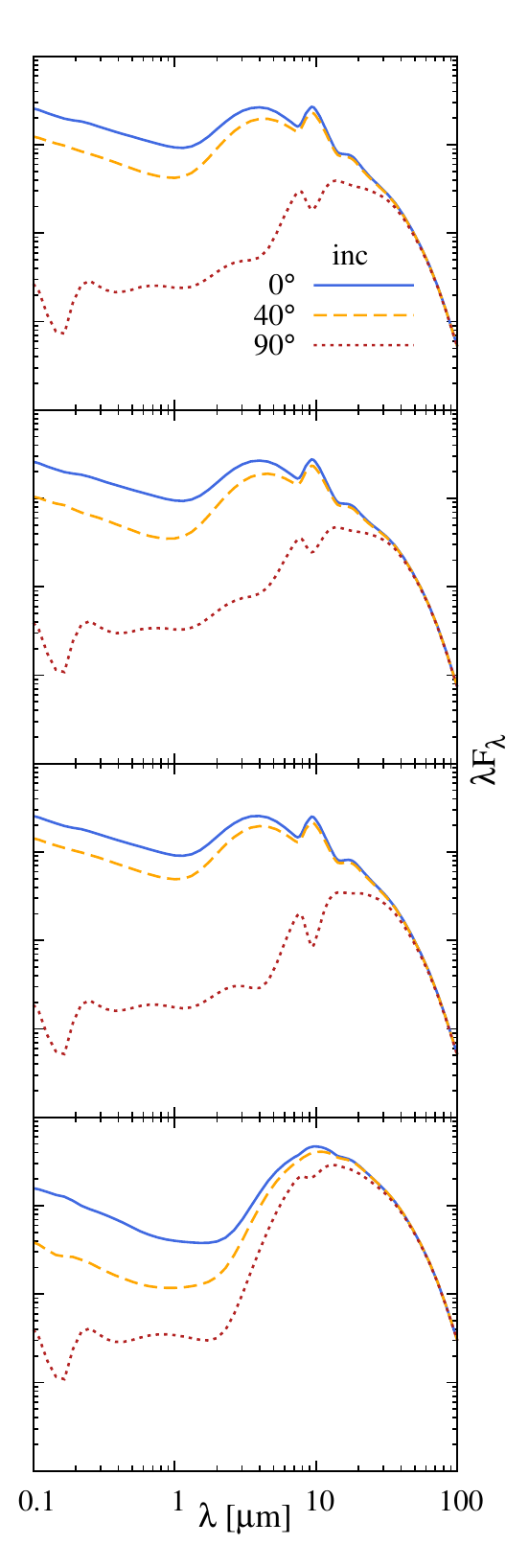}
\caption{Model images obtained with radiative transfer simulations for the four geometries shown in Fig.~{\ref{fig:geoscheme}}, each at three different viewing angles, and their corresponding SEDs. Inclinations $0^{\circ}$ and $90^{\circ}$ respectively correspond to face- and edge-on views, while the $i=40^{\circ}$ viewing angle provides a view through the shell of the polar component. In the examples here we display the images at $\lambda = 36\,\micron$ as at these wavelengths they most clearly show the difference in morphology of the models while being displayed in the logarithmic color scale with the same lower and upper cuts.}
\label{fig:modImgSED}
\end{figure*}

Fig.~{\ref{fig:modImgSED}} showcases $\lambda = 36\,\micron$ radiative transfer images and SEDs of the four different model geometries described in \S \ref{sec:mod-geo} and sketched in Fig.~{\ref{fig:geoscheme}. Viewed face-on, all the models appear circular, despite the emission from the accretion disk being tilted with respect to the system axis. The reason is that, in this viewing angle, the dust structure is optically thin and the regions of enhanced emission from the cone and counter-cone compensate each other. Inside the paraboloid shell model, a dark ring is visible. This is simply a consequence of the geometry: the accretion disk radiation emitted in the direction of the inner part of the paraboloid walls is absorbed by the dusty disk. The sphere shell model appears less extended than the others, as it is the only model in which the dust fills the entire polar region, and not only the walls of the cone. In the intermediate viewing angle which is going through or grazing the cone walls, all the model images feature similar, roughly hourglass-shaped structures. The hourglass shape is most prominent in the paraboloid shell model, again due to its particular geometry with curved walls. At edge-on inclinations, the three cone-like models exhibit a prominent bar-like structure, due to the anisotropic emission of the tilted accretion disk which is illuminating only one side of the wall. The polar emission region is much wider in the sphere shell model. The surface of a hyperboloid asymptotically approaches the cone surface at larger distances, and thus they appear almost identical; they are significantly different only at the inner few parsecs, which are unresolved in the VISIR images. The SEDs of all the cone-like models are similar at all viewing angles. The paraboloid shell model has a deeper silicate feature which is due to the missing flux from the inner region of its walls which are not directly illuminated. The SEDs of the torus$+$sphere shell model are much more isotropic, as expected from a model featuring a radially symmetric component.

In this particular model realization ($\Delta_{disk}=20^{\circ}$), the silicate absorption feature would turn to emission for $i<70^{\circ}$ (if there is no foreground extinction screen). This corresponds to the dust covering factor of 0.34, which is lower than the average value inferred from the ratio of type 1 to type 2 sources. However, both the disk and the polar components are thick at optical wavelengths. Thus, in optical classification, the covering factor would be 0.5, which is in agreement with the covering factor distribution inferred by recent IR studies using large samples \citep{Netzer2016, Stalevski2016}.

By inspecting Fig.~{\ref{fig:modImgSED}} it seems that all four models have potential to explain the bar seen in VISIR images. However, for a proper comparison, we need to simulate how would model images appear when observed with VISIR. We achieve this by carrying out the following procedure for the model images: (a) rotation to match the on-sky orientation of Circinus, (b) resampling to match the VISIR pixel size, (c) convolution with the observed PSF, and (d) inclusion of the same level of noise as estimated from the observed images. The resulting simulated observations of the model images are presented in Figs.~{\ref{fig:cirhyp}, \ref{fig:cirpar} and \ref{fig:cirsph}}, for the models consisting of disk$+$hyperboloid, disk$+$paraboloid and torus$+$sphere shell, respectively. Each includes the cases of an aligned and a tilted accretion disk. The case of the disk$+$cone shell model is omitted, as it is virtually the same as the hyperboloid model. We find that all the models with the accretion disk aligned with the system axis are clearly not consistent with the observations as they feature a distinct X-shaped morphology.

On the other hand, for a tilted accretion disk, cone-like models exhibit a bar-like feature, similar to the bar seen in the VISIR images. It is evident that the disk$+$hyperboloid model offers a qualitatively good match with observed morphology, featuring a straight bar with the same extension and orientation. The paraboloid model resembles the observations not so well on the account of showing a warp in the bar and a changing orientation at different wavelengths, due to its curved walls, which is not observed.

Finally, the model of the clumpy torus surrounded by a sphere shell can be ruled out as its morphology does not provide a good match to the observations at any wavelength: the bar-like feature appears prominently only in the $18\,\micron$ image, but it is too wide and significantly misaligned with respect to the orientation of the observed bar. If the model is rotated to match the orientation of the bar, then (a) the opening angle of the torus would not match the orientation of the observed ionization cone and (b) the accretion disk in the model would not match the orientation of the inner part of the observed maser disk. We note that the VISIR images cannot constrain the dust structure in the unresolved parsec-scale region, i.e., whether the compact structure is a thin disk or a thick torus. However, we can exclude that the observed MIR bar is merely a consequence of the collimated illumination of uniformly distributed ambient dust.

\begin{figure*}
\centering
\includegraphics[width=1.0\textwidth]{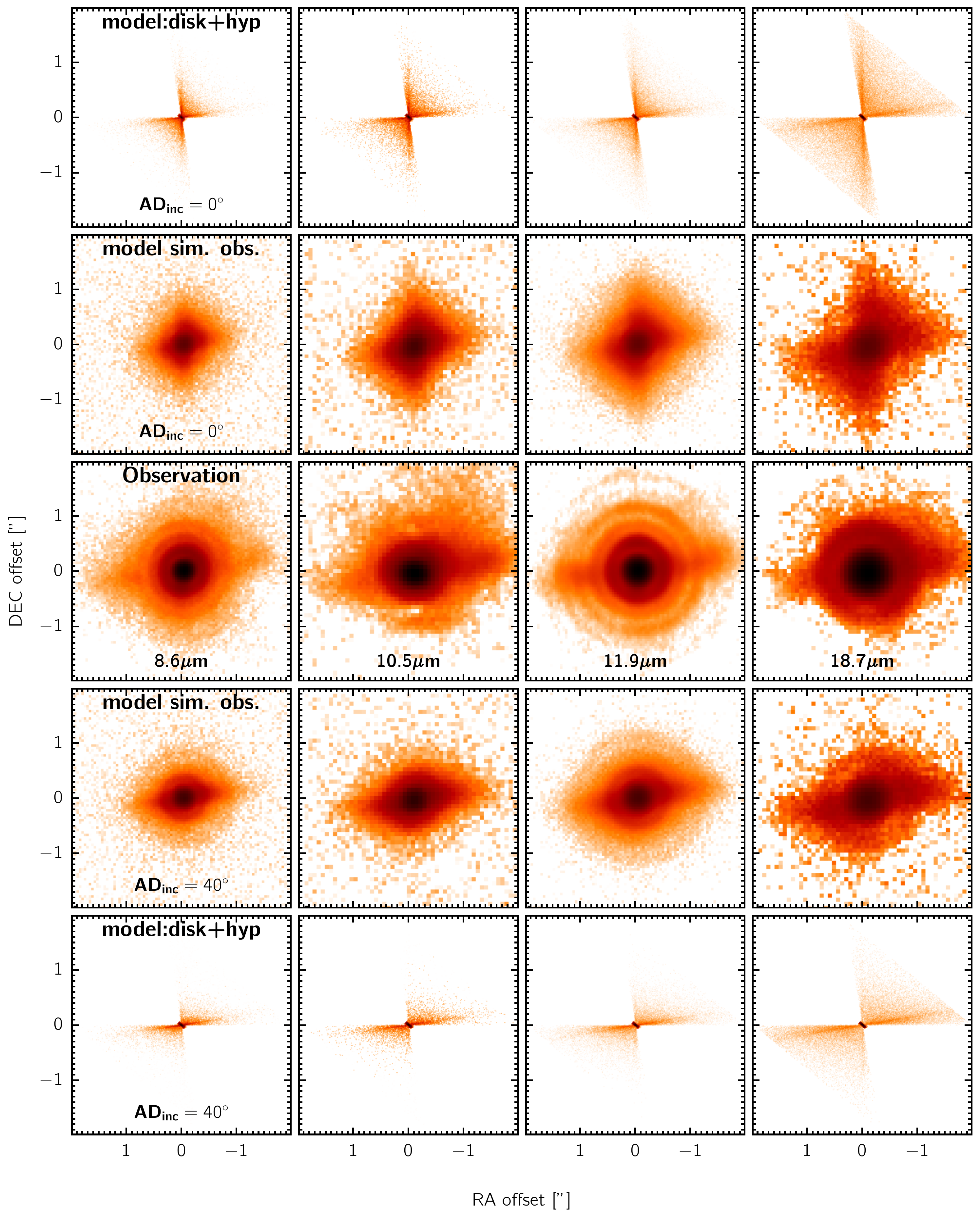}
\caption{The disk$+$hyperboloid model as it would appear when observed with VISIR, compared to the actual VISIR images. The model images are rotated to match the Circinus on-sky orientation [rows (1) and (5)]. Then they are resampled to the VISIR pixel size and convolved with corresponding observed PSF, including the same level of noise as estimated from the observed images [rows (2) and (4)]. The two top rows show the model with the accretion disk aligned with the system axis, while the two bottom rows show the model with the disk tilted by $40^{\circ}$. To facilitate comparison, the middle (third) row repeats the observed images shown in Fig.~\ref{fig:cirVLT}. Columns feature images at $8.6$, $10.5$, $11.9$ and $18.7\,\micron$, from left to right. It is evident that the model with a tilted disk provides a qualitatively good match with the observations, unlike the aligned disk model which features an X-shaped morphology.}
\label{fig:cirhyp}
\end{figure*}
\begin{figure*}
\centering
\includegraphics[width=1.0\textwidth]{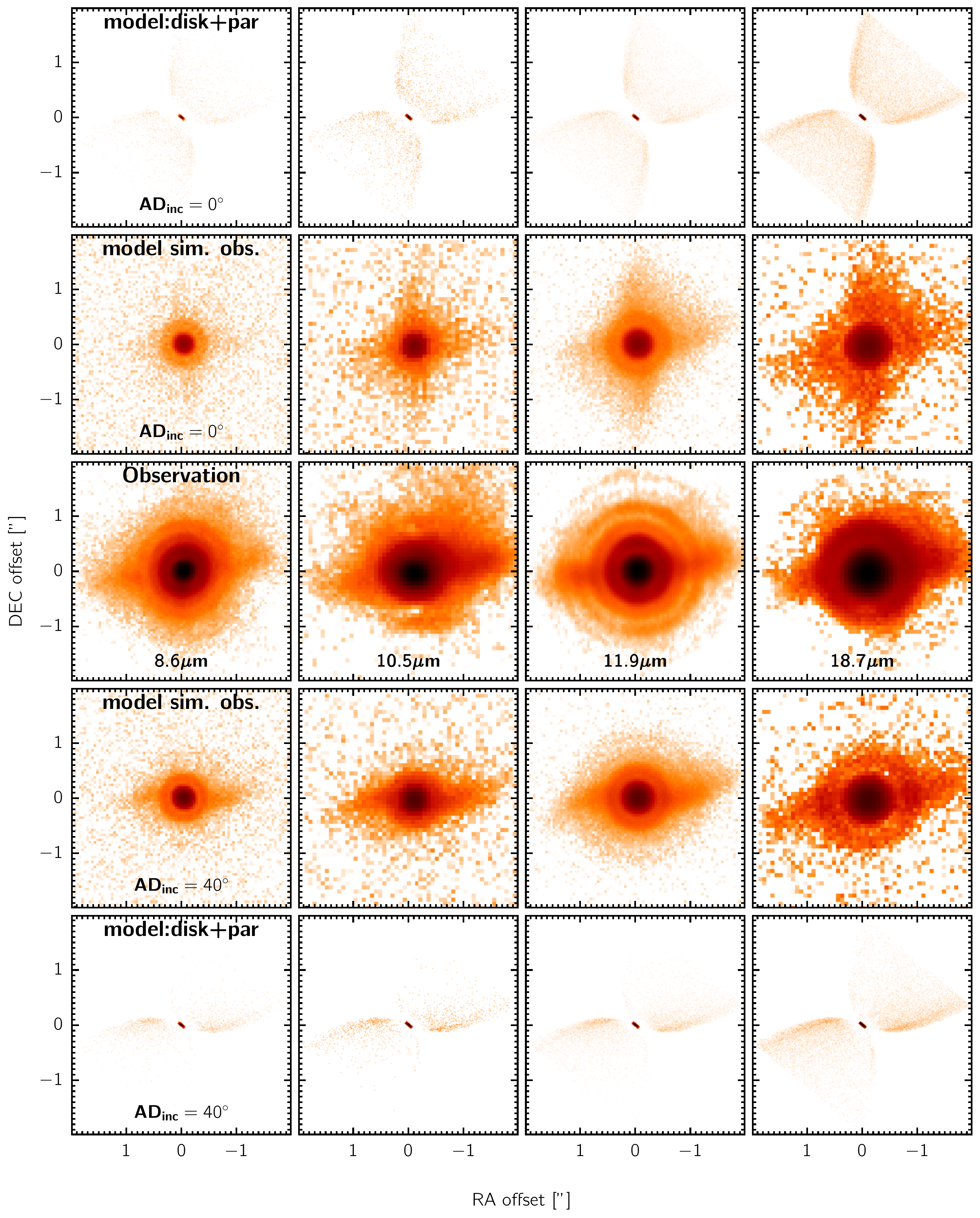}
\caption{The same as in Fig.~\ref{fig:cirhyp} but for the disk$+$paraboloid shell model. As in the hyperboloid shell case, aligned and tilted accretion disk models results in X-shaped and bar-like morphology, respectively. However, the paraboloid shell model exhibits a notable warp in the bar and a change of the bar orientation as a function of the wavelength.}
\label{fig:cirpar}
\end{figure*}
\begin{figure*}
\centering
\includegraphics[width=1.0\textwidth]{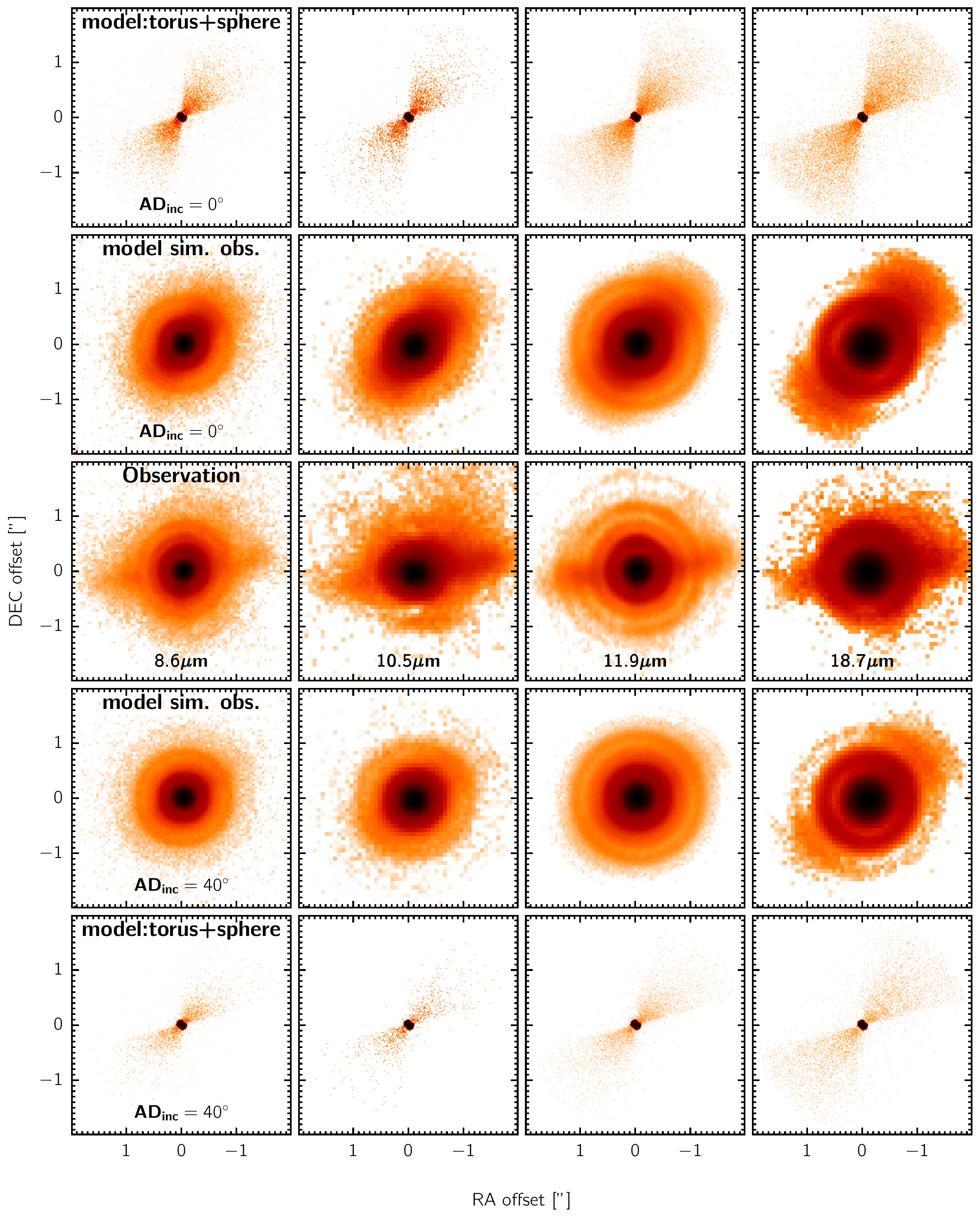}
\caption{The same as in Fig.~\ref{fig:cirhyp} but for the torus$+$sphere shell model. Only the longest wavelength image features a prominent bar-like morphology, however too wide and with orientation inconsistent with the observed one. If the model is rotated to match the orientation of the bar, then the opening angle of the torus would not match the orientation of the observed ionization cone. At the other three wavelengths, the model images are dominated by the PSF and show no structure resembling the observed bar.}
\label{fig:cirsph}
\end{figure*}

\subsection{Model vs. data: SED}
\label{sec:res-SED}

\begin{figure*}
\centering
\includegraphics[width=1\textwidth]{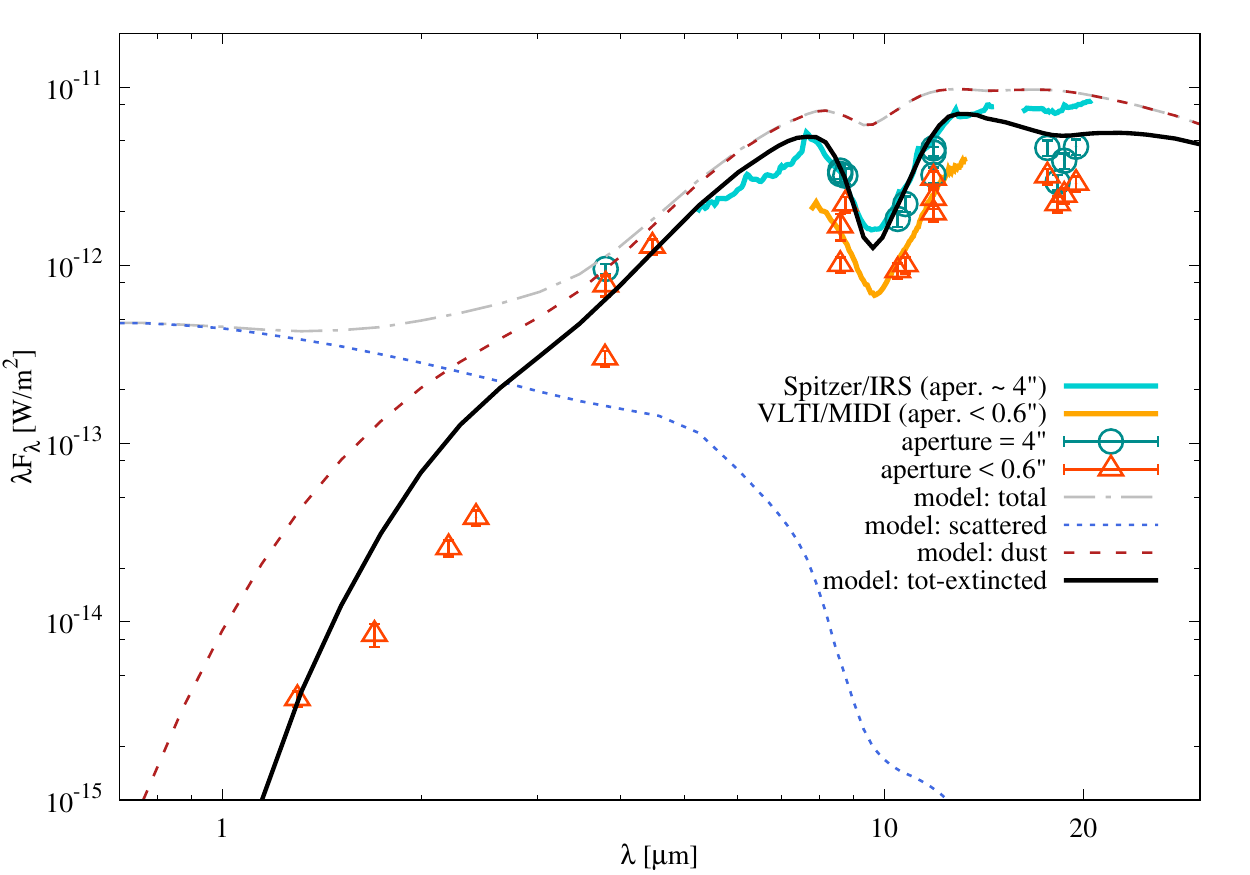}
\caption{Comparison of the observed SED compiled from the literature with one representative model SED. Large-scale aperture photometry ($4\arcsec$) is shown in green circles, while photometry extracted from apertures comparable or smaller than the resolution limit of VISIR ($0.6\arcsec$) is marked by red up-pointing triangles. The measurement uncertainties are smaller or comparable to the plotted symbol size. The aperture of the Spitzer/IRS spectrum from \citet{Asmus2014} is comparable to the total aperture of VISIR in the $5.2-14.5\,\micron$ range, while significantly larger at longer wavelengths. The MIDI spectrum from \citet{Tristram2014} was extracted using a $\sim0.54\arcsec\times0.52\arcsec$ aperture and hence corresponds to the unresolved nucleus with VISIR. The model SED is decomposed into its total (dash-dotted gray), scattered (dotted blue) and dust (dashed dark red) components. For a realistic comparison, foreground absorption must be taken into account: the total model flux extincted by foreground screen of average $\tau_{9.7}=1.6$ is shown in solid black line.}
\label{fig:sed_vs_mod}
\end{figure*}

Since the disk$+$hyperboloid model provides the best match to the observed MIR morphology, we proceed with the more detailed comparison using this model only. Because both the disk$+$hyperboloid and the disk$+$cone shell model appear virtually the same at larger distances and cannot be distinguished by current data, everything said about the hyperboloid model also applies to the cone shell model. The difference between the two geometries becomes relevant only when one is able to resolve the inner few parsecs. The dust structures on parsec scales are the subject of a dedicated investigation by modeling the interferometric MIDI data which will be presented in an accompanying paper (Stalevski et al., in prep.).

We generated a grid of disk$+$hyperboloid model SEDs for different values of the disk parameters $R_{\text{out}}$ (outer radius), $\tau_{\text{9.7}}$ (edge-on optical depth), $\Delta$ (disk flaring angle), $p$, $q$ (density law parameters), $L_{\text{AGN}}$ (accretion disk luminosity) and compared them to the SED and spectra of Circinus compiled from the literature (see. \S \ref{sec:obs-lit} and table \ref{tab:data}). We compare simply the monochromatic flux of the model which is closest to the central wavelength of the filter. The spectral resolution of the model is comparable to the filter band width and thus synthetic photometry is not feasible, nor necessary for our qualitative comparison. For an informative comparison, we plot photometric data points with different symbols depending on whether they were obtained with an aperture comparable to the resolution limit of VISIR ($\leq 0.6\arcsec$) or have significantly larger apertures rather corresponding to the total flux in the VISIR images ($4\arcsec$). 
The size of the model simulation box corresponds to the total field of view of VISIR, and thus, should be compared to the $4\arcsec$ data. The small aperture data provide only lower limits, especially in the NIR where no data is available with apertures comparable to the total VISIR field of view. The Spitzer/IRS spectrum (taken in 2004) is rescaled to match the total aperture VISIR photometry at $8.6\,\micron$, to account for the increase of Circinus flux between 2008 and 2010 \citep{Tristram2014}.

Before proceeding with the SED comparison it is important to note that there is evidence for a significant amount of foreground absorption that must be taken into account. Employing high $8$-$13\,\micron$ spatial resolution spectroscopy with T-ReCS on Gemini-South, \citet{Roche2006} found that the extended emission East from the nucleus displays a much deeper silicate absorption than on the West side \citep[similar variations in the absorption profile of $3.4\,\micron$ feature have been found by][]{Colling2009}. They attribute this to the extinction due to the inclined host galaxy disk being in the foreground, extending over tens of parsecs. They derive the amount of extinction by fitting the profiles of the silicate features extracted for different slit positions along the extended bar and in the perpendicular direction. These values represent a lower limit as they are obtained assuming a featureless black body continuum (no silicate emission or absorption) for the source emission. This is consistent with our model, which assumes only graphite in the polar region. If the model of the extended dust would include silicates, higher levels of foreground absorption would be inferred, as also discussed by \citet{Roche2006}. Thus, our results do not rule out the presence of silicates in the polar region. Here, we use a simple approximation of this foreground absorption by generating an extinction map with a linear gradient mostly in the East-West direction. For the absorption profile we use the extinction curve based on Galactic interstellar dust from \citet{Tristram2007}. The gradient is obtained by fitting the absorption values extracted from the spectra at $1.6\arcsec$ distance from the nucleus in all four directions. This leads to a nuclear foreground extinction of $\tau_{\text{9.7}}=1.6$ consistent with the observations.

A number of different parameter combinations result in model SEDs qualitatively consistent with the data. In Fig.~{\ref{fig:sed_vs_mod}}, we show one representative SED decomposed into its scattered and dust emission components, and attenuated by dust of the host galaxy. We see that scattered light has a significant contribution shortward of $3\,\micron$; however, it is completely extinguished by the foreground dust screen. The model SED has a weak silicate absorption feature, which becomes much deeper and matches the data after the foreground extinction screen is applied. We note that there are models featuring silicate absorption profiles that could fit the data with much less or without the foreground extinction screen. However, foreground absorption is robustly established by the very deep off-center optical depth values \citep{Roche2006}, and we consider realistic only those models which provide a good match including the inferred amount of extinction.

In Fig.~\ref{fig:BestModImg}, we compare the VISIR images of Circinus with the simulated observations of our representative model, including background noise and foreground extinction. Since the observations are diffraction-limited, we include in the plot the instrumental PSF, approximated by an azimuthally symmetric structure of the reference star, to allow easier interpretation of the images. We see that the simulated model images provide a good match to the observed morphology in all filters. 

\begin{figure*}
\centering
\includegraphics[width=1\textwidth]{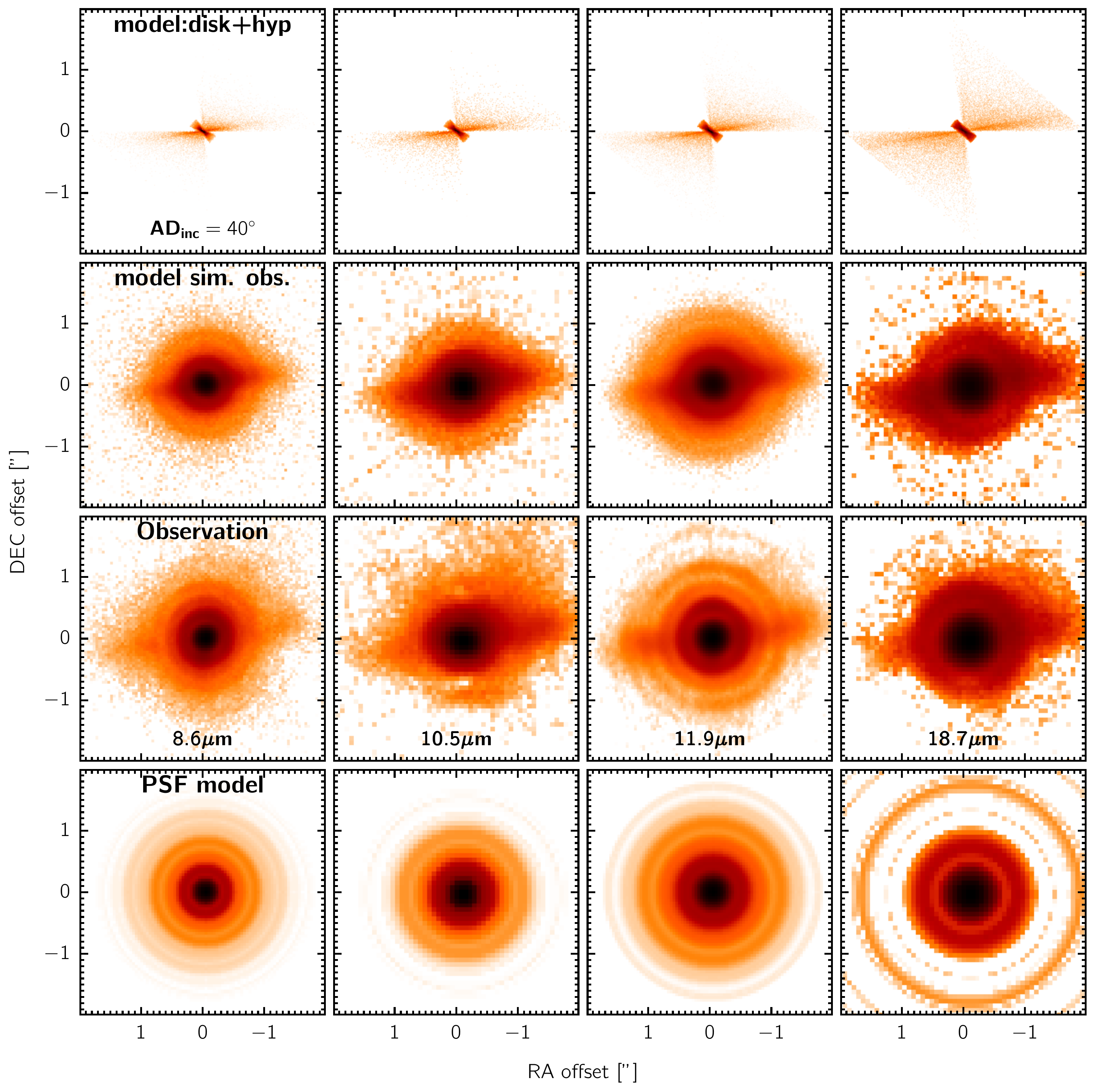}
\caption{Comparison of the VISIR images of Circinus with the representative model (whose SED is shown in Fig.~\ref{fig:sed_vs_mod}), including foreground extinction. From top to bottom, the rows show: (1) the model images; (2) the model images as they would appear when observed with VISIR; (3) the images of Circinus acquired with VISIR; (4) our approximation of the observed PSF.}
\label{fig:BestModImg}
\end{figure*}

\begin{figure*}
\centering
\includegraphics[width=1\textwidth]{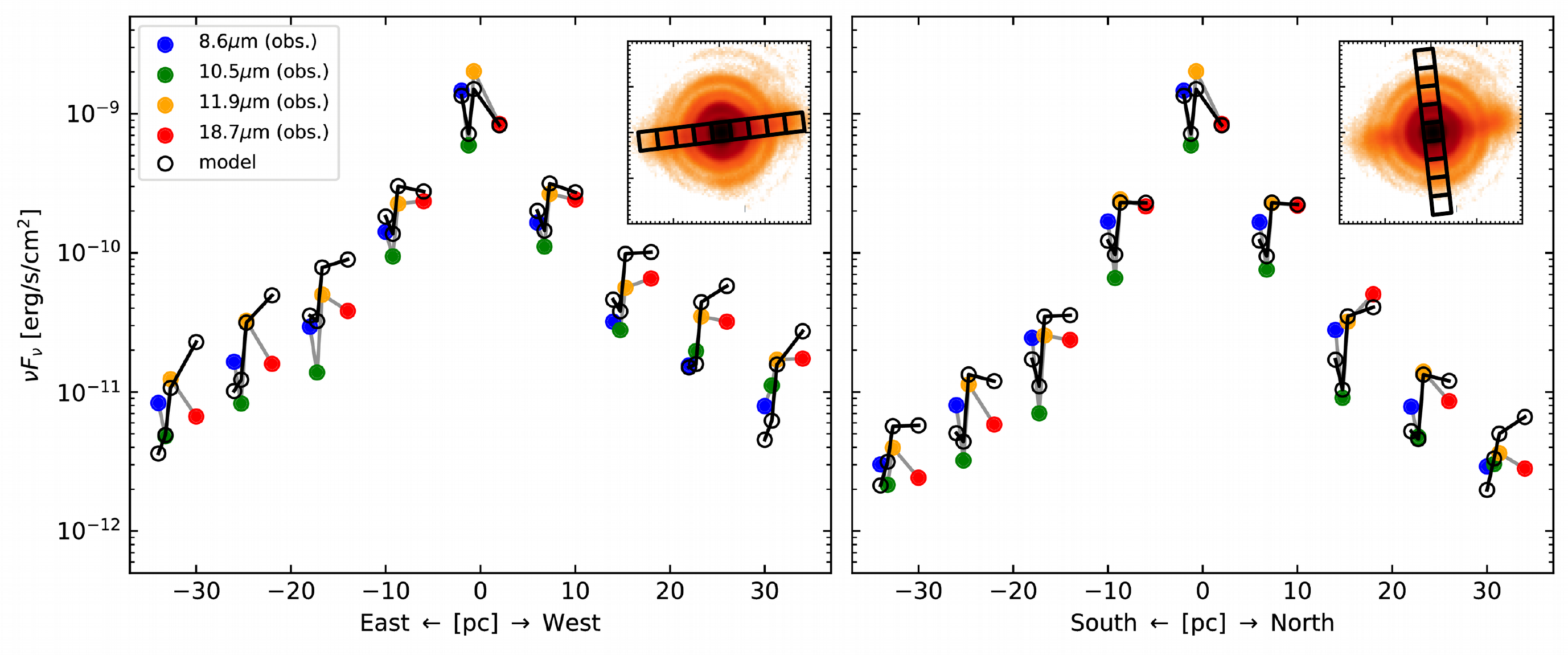}
\caption{Comparison of observed and model photometry (from images shown in Fig.~\ref{fig:BestModImg}) extracted in $0.4\arcsec\times0.4\arcsec$ aperture fields in the directions along the bar (left) and perpendicular to it (right), as indicated in the inset plots. The central position corresponds to the unresolved nucleus.}
\label{fig:mini-SED}
\end{figure*}

We compare the VISIR images and the simulated observations of the model in more detail in Fig.~\ref{fig:mini-SED} by measuring fluxes at different positions extracted from $0.4\arcsec\times0.4\arcsec$ apertures along the MIR bar and perpendicular to it, similar to the spectral extraction by \citet{Roche2006}. An overall qualitatively good agreement is evident, especially at the central position and close to it. At positions further away from the center, the model overpredicts the $18\,\micron$ flux. There are several possible causes of this mismatch: (a) an inhomogeneous (and possibly steeper) foreground absorption screen (which in our model is assumed to be continuous with linear decrease in the East-West direction); (b) an additional heating mechanism contributing to the dust heating (e.g. collisional dust heating; line emission heating in the coronal emission zone; photon trapping); or (c) a different extinction curve than here assumed based on Galactic ISM. At the farthest positions ($\sim30\,$pc), the measurements are dominated by noise, while simulated observations of the model depend highly on the method used to approximate the PSF.

All the presented comparison and analysis is leading us to conclude that the here presented model consisting of a compact dusty disk and large scale dusty hyperboloid/cone shell illuminated by a tilted accretion disk with anisotropic emission pattern plausibly represents the actual dust structure in Circinus.

\subsection{The origin of the dust in the polar region}
\label{sec:origin}

Our model is purely phenomenological, designed to reproduce the MIR observations of Circinus, but with a solid base on a number of observations across a range of wavelengths and of different spatial scales. However, the presented model is supported by a growing number of theoretical works that employ different methods to investigate dusty winds and outflows in AGN. 
Using a photoionization model of a dusty gas, \citet{Netzer-Laor1993} were the first to suggest that a number of unexplained features in the relation between the narrow- and broad-line emitting regions could be resolved naturally if dust is embedded in the narrow line-emitting gas. \citet{Konigl-Kartje1994} considered the implications of centrifugally driven, self-similar winds from the surfaces of accretion disks, incorporating the effect of radiation pressure in an approximate way. They found that disk-driven hydro-magnetic winds could be the origin of bipolar jets and outflows, and that the obscuring dusty tori represent outer regions of the disk wind. Building up on the work of \citet{Konigl-Kartje1994} and expanding the explored parameter space, \citet{Keating2012} and \citet{Gallagher2015} investigated further dusty disk winds, driven by both magnetocentrifugal forces and radiation pressure as an origin of the torus and outflows. \citet{Roth2012} used Monte Carlo radiative transfer calculations to determine the radiation force on a dusty gas within a few tens of parsecs and reach similar conclusions as their predecessors: radiation pressure drives away the gas and dust to the polar regions, leaving behind what may constitute the obscuring dusty torus. Using three-dimensional hydrodynamic simulations including X-ray heating and radiation pressure, \citet{Wada2012} found that radiation feedback drives a vertical circulation of gas (a ``fountain'') in the central few to tens of parsecs, creating a turbulent, geometrically thick obscuring region. In a series of works \citep{Dorodnitsyn2011, Dorodnitsyn2012, Dorodnitsyn-Kallman2012, Dorodnitsyn2016}, IR radiation is incorporated in radiation-hydrodynamics simulations adopting a flux-limited diffusion approximation. These works find that the obscuration is produced at parsec scales by a dusty wind which is supported by infrared radiation pressure on dust grains. Another incarnation of radiative magnetohydrodynamics simulations is presented by \cite{Chan-Krolik2016, Chan-Krolik2017} who solve the magnetohydrodynamics equations simultaneously with the IR and UV radiative transfer equations. They find that the outflow launched from the torus inner edge by UV radiation expands in solid angle as it ascends, while IR radiation continues to drive the wide-angle outflow further away.

Most of the works discussed above focus on the sub-parsec to parsec scale, where wind is launched, but do not trace it to few tens of parsecs, (which is the scale of interest for Circinus), or they provide only general predictions for the wind properties. Due to this, a direct comparison to our model is not possible except in a few cases and only in a very general way. 
For example, \citet{Gallagher2015} predict hour-glass shape for the polar dusty winds considered in their framework and provide a MIR radiative transfer image of their solution; this image looks remarkably similar to the model images of the cone-like structures we assume in this work.
\citet{Wada2016} implemented supernova feedback in the ``radiation-driven fountain'' of \citet{Wada2012} and produced a model with a black hole mass similar to that of Circinus. They found that polar outflows in a shape of a double hollow cone are produced assuming a supernova rate about factor of ten larger than observed for Circinus. However, this could be a consequence of certain assumptions than do not allow the numerical simulations to capture the full complexity of the environment around SMBHs (for example, neglecting the self-gravity of the gas or insufficient resolution) and result in lower supernova efficiency than in reality. 
Finally, \cite{Honig2017} presented a library of IR emission based on an empirical disk$+$wind model and showed that such a model can account for the polar elongation in the case of NGC 3783. The geometry of their model has a very similar setup to ours; however, the necessary number of models we needed to calculate for the purpose of our work is modest compared to their library and does not allow further comparison in a meaningful way.

In summary, while all the discussed works are not without their shortcomings due to the particular assumptions and approximations employed, a solid theoretical background has emerged, showing that: (a) radiatively-driven winds are a plausible mechanism of bringing the dust to the polar regions of AGN, and (b) the polar winds are expected to take shape of a double cone-like structure.

\section{Summary and conclusions}
\label{sec:sum}

Recent findings of significant MIR emission in the polar regions of local AGN challenge the widely accepted picture in which the parsec-sized ``torus'' is responsible for the entire AGN dust emission and may demand a change of paradigm. One of the sources showing clear polar extended dust is the archetypal type 2 AGN in the Circinus galaxy. Up-to-date highest quality images obtained with the upgraded VLT/VISIR instrument feature a prominent bar extending in the polar direction of the system. In this work, we presented a phenomenological model for the dust emitting regions in this source consisting of a compact dusty disk and a large scale dusty hollow cone-like region illuminated by a tilted accretion disk with anisotropic emission. The model is supported by observations across a range of different wavelengths and spatial scales. Using Monte Carlo radiative transfer simulations we produced a grid of images and SEDs for different geometries that have a potential to explain the observations: a dusty disk$+$dusty polar region in the form of a cone shell, a hyperboloid shell and a paraboloid shell. In addition, we tested a model in which a clumpy dusty torus is collimating the accretion disk radiation which then illuminates the diffuse host galaxy dust that surrounds the torus. Based on the comparison with the observed MIR morphology and the measured SED we conclude that:

\begin{itemize}
  \item The model in which the dusty torus collimates the accretion disk emission illuminating the surrounding ambient dust fails to reproduce the observed MIR morphology, as it does not feature the prominent polar bar.
  
  \item The model in which the walls of the polar dust shell are in the shape of a paraboloid surface do not match the observed structures well because the model images feature a warp and a change of the position angle of the bar, which are both not observed.

  \item Models of the polar dust in the form of a hollow cone or a hyperboloid shell (which are both effectively the same at large distances) are able to reproduce the observed MIR morphology at all wavelengths. Furthermore, such models provide a good match to the entire IR SED of Circinus, including resolved photometry extracted from apertures at different positions along the polar bar and perpendicular to it.
\end{itemize}

\emph{We conclude that our model consisting of a compact dusty disk with a large scale dusty hyperboloid/cone shell is plausibly a good representation of the dust structure of the AGN in the Circinus galaxy.}

Our results call for caution when attributing thermal dust emission of unresolved sources entirely to the torus and warrant further investigation, including detailed modeling of other sources showing polar elongation of their MIR emission as well as developing a theoretical framework that would explain the origin of the dust in the polar regions. Our model of the AGN in Circinus can be used as a prototype and as a guideline for such studies.

\section*{Acknowledgments}

We thank the referee for useful suggestions; Sebastian H\"{o}nig for many illuminating discussions; Claudio Ricci for careful reading and comments on the manuscript. MS acknowledges support by the Ministry of Education, Science and Technological Development of the Republic of Serbia through the projects Astrophysical Spectroscopy of Extragalactic Objects (176001) and Gravitation and the Large Scale Structure of the Universe (176003); by FONDECYT through grant No.\ 3140518; by the Center of Excellence in Astrophysics and Associated Technologies.  This research made use of Astropy, a community-developed core Python package for Astronomy \citep{Astropy2013}.


\bibliographystyle{mnras}
\input{circinus-I_rev_28-08-2017_astroph_v2.bbl}

\appendix

\bsp

\label{lastpage}

\end{document}